\documentclass[review]{elsarticle}
\usepackage{amsmath}
\usepackage{amssymb}
\usepackage{graphicx}
\usepackage{color}
\graphicspath{{images/}}
\journal{Superlattices and Microstructures}

\begin{document}

\title{Non-equilibrium condensation in periodic polariton lattices}
\author{Berihu Teklu}
\author{Dmitry Solnyshkov}
\author{Guillaume Malpuech\corref{mycorrespondingauthor}}
\cortext[mycorrespondingauthor]{Corresponding author}
\ead{malpuech@univ-bpclermont.fr}
\address{Institut Pascal, PHOTON-N2, Universit\'e Clermont Auvergne, University Blaise Pascal/CNRS,4 avenue Blaise Pascal, 63177 Aubi\`ere Cedex, France}

\begin{abstract}
We study theoretically the Bose-Einstein condensation in periodic polariton lattices in presence of a localized reservoir. Using the Schr\"odinger and Boltzmann equations, we find analytically the parameters determining the choice of the bandgap for condensation on the defect states.
\end{abstract}

\maketitle

Bose-Einstein condensation \cite{Pitaevskii} is a vast field of research, and condensates in periodic lattices have become an important topic in this field \cite{Bloch2005}. Many important results have already been obtained for atomic condensates in the lattices, which extend the possibilities for the modification of the condensate's characteristics \cite{Bloch2008,Morsch2006,Ritsch2013}. At the same time, condensates of various quasi-particles demonstrating original properties are also studied very actively \cite{Demokritov2006,Vainio2015,Eisenstein2014,Misochko2004}. Among them, exciton-polaritons and polariton condensates \cite{Kasprzak2006} are rapidly developing thanks to their important advantages. Polaritons, which are half-light half-matter quasiparticles obtained in semiconductor microcavities in the regime of strong light-matter coupling \cite{Microcavities}, combine the properties of light, such as high propagation velocities and the possibility of complete measurement of the wavefunction, with the properties of matter, providing strong interactions. Due to the finite lifetime, polariton condensation can be an out-of-equilibrium process \cite{Kasprzak2008,Carusotto2013rev}, which allows considering theoretically and obtaining experimentally polariton condensates which are not in the system's ground state.

Recent technological advances have allowed to create periodic potentials for polaritons using metal deposition\cite{Yamamoto2013}, surface acoustic waves \cite{Cerda2010}, and etching \cite{Ferrier2011}. The potentials can be created by optical means as well \cite{Dreismann2014}. The technical possibilities for potential engineering become very large, including even non-periodic potentials with fractal spectrum \cite{Tanese2014}.
 Already the first experiments in periodic lattices have demonstrated that the periodic potential strongly affects the condensation by creating negative mass states, which become preferential for polariton condensation \cite{Tanese}. Indeed, the experimental configuration often includes a non-resonant localized optical pump, which creates electrons and holes forming excitons in a small-sized spot (which can be from 2 to 20 microns\cite{Jacqmin2014}). Exciton-phonon and exciton-exciton interactions then lead to the relaxation of polaritons towards the ground state. However, the reservoir also creates a potential for polaritons by exciton-exciton interactions, which the polaritons feel via their excitonic fraction. This potential repels polaritons occupying positive mass states, but attracts polaritons in negative mass states, which therefore concentrate directly under the reservoir. The inclusion of the periodic potential providing the negative mass can therefore lead to the condensation on such states, in particular, on the defect states formed in the gaps separating the allowed energy bands. This has already been demonstrated experimentally, together with the evolution of a defect state into a gap soliton \cite{Tanese}, where the interactions within the condensate start to dominate the potential created by the reservoir. Negative mass, either inherent to the particles or induced by the lattice potential, is in general favorable for the observation of self-localized condensates: bright solitons were also observed for polaritons in the non-parabolic part of their dispersion without any lattices \cite{Sich2011}. Gap solitons have also been studied in atomic condensates in optical lattices \cite{Eiermann2004}.

However, the problem of the condensation on the defect states in periodic polariton lattices has never been studied analytically. Depending on the potential amplitude, many allowed bands and many bandgaps can appear, but in which particular gap the condensation occurs is an important question which was never answered in general. A recent experimental study demonstrated switching between the different gap states driven by the conditions of a particular experiment \cite{Amo2015}; another study has demonstrated the formation of a gap soliton in the second bandgap \cite{Tanese}, but why did it occur precisely there, has never been explained. In both of these works, as well as in other works devoted to self-localized condensates\cite{Ostrovskaya2013}, the theoretical description was based on numerical solution of different types of Schr\"odinger equation \cite{Ge2013}. At the same time, a powerful Boltzmann-Gross-Pitaevskii approach for the description of dynamical condensation on quantized states has been developed in series of previous works of polaritons \cite{Ferrier2011,Galbiati2012,Kalevich2015,Sturm2014}, but it has been used only numerically so far.

In this work, we present an analytical study of a general case of polariton condensation a 1D periodic lattice, based on coupled Schr\"odinger and Boltzmann equations \cite{Galbiati2012} using the tight-binding and the effective mass approximations, taking into account the polariton lifetime, overlap integrals and temperature. We find the dependence of the preferential bandgap for the condensation on the system's parameters and demonstrate that the system can be driven either to condense always on the lowest bandgap, or to the higher bandgaps. Condensation in the true ground state is also possible, but requires specific conditions.

\section{Model}\label{s:mod}
Our analytical model consists of several levels. We begin with the description of cavity exciton-polaritons, then calculate the parameters of the tight-binding description of a perodically modulated polariton wire, then calculate the tight-binding dispersion and effective mass, then solve the Schrodinger equation to find the shape of the defect states created by the reservoir potential in the gaps, and then, finally, describe the condensation on these defect states by using the Boltzmann equations.

Microcavity polaritons are quasiparticles resulting from a superposition of cavity photons and quantum well excitons \cite{Weisbuch}. The strong exciton-photon coupling can be described using a model of two coupled quantum oscillators, with the Hamiltonian of the system written as \cite{Microcavities,CavityPol}
\begin{equation}
H=E_{ph}\hat a_{k}^{\dagger} a_{k}+E_{ex}\hat b_{k}^{\dagger}b_{k}+\hbar\Omega(\hat a_{k}^{\dagger} b_{k}+\hat b_{k}^{\dagger} \hat a_{k})
\end{equation}
where $\hat a_{k}^{\dagger}$ and $\hat b_{k}^{\dagger}$ are the photon and exciton creation operators. The Rabi splitting $\hbar\Omega$ determines the strength of dipole interaction between the exciton and the light field. 
The above Hamiltonian is diagonalized by   
\begin{align}
\hat P_{k}&=X_{k}\hat b_{k}+C_{k}\hat a_{k},\\
\hat Q_{k}&=-C_{k}\hat b_{k}+X_{k}\hat a_{k}
\end{align}
Then the effective polariton Hamiltonian becomes
\begin{equation}
\hat H_{p}=E_{LP}(k)\hat P_{k}^{\dagger}\hat P_{k}+E_{UP}(k)\hat Q_{k}^{\dagger}\hat Q_{k}.
\end{equation}
Here, ($\hat P_{k}$,$\hat P_{k}^{\dagger}$) and ($\hat Q_{k}$,$\hat Q_{k}^{\dagger}$) are the eigenmodes of the system. The
eigenvectors of this Hamiltonian are a superposition of the exciton and photon states which can be represented as
\begin{align}
|UP\rangle&=C_{k}|ex\rangle+X_{k}|ph\rangle\\
|LP\rangle&=X_{k}|ex\rangle-C_{k}|ph\rangle
\end{align}
where 
\begin{align}
|X_{k}|^2&=\frac{1}{2}\left(1+\frac{\Delta E(k)}{\sqrt{\Delta E(k)^2+\hbar^2\Omega^2}}\right),\\
|C_{k}|^2&=\frac{1}{2}\left(1-\frac{\Delta E(k)}{\sqrt{\Delta E(k)^2+\hbar^2\Omega^2}}\right),
\label{eq:HF}
\end{align}
are the standard Hopfield coefficients \cite{Hopfield,Sermage} describing the fraction of the exciton and the photon content of the polariton. 
The quantity $\Delta E$
\begin{equation}
\Delta E(k)=E_{ph}(k)-E_{ex}(k)+\eta
\end{equation} 
is the difference between the cavity photon resonance energy and the quantum well exciton energy  and $\eta$ is the exciton-photon detuning.
The two coupled mode eigenstates of the system are called the upper polariton (UP) and the lower polariton (LP), corresponding to the higher and the lower energy states, respectively. The dispersion of the lower polariton branch is given by
\begin{equation}
E_{LP}(k)=\frac{1}{2}\left[(E_{ex}+E_{ph})-\sqrt{(E_{ex}-E_{ph})+\hbar^2\Omega^2}\right].
\end{equation}

In this work, we consider only polaritons from the lower polariton branch. Since the dispersion is non-parabolic, the polariton mass depends on the wavevector. Another important parameter depending on the wavevector is the polariton lifetime. Quantitatively, the lifetime of polaritons is directly determined by the fraction and lifetime of each individual photon and exciton component, according to

\begin{equation}
\tau=\left(\frac{|X_{k}|^2}{\tau_{C}}+\frac{|C_{k}|^2}{\tau_{X}}\right)^{-1},
\end{equation}  
where $\tau_{C}(\tau_{X})$ are the bare photon (exciton) lifetime. The excitonic and photonic fractions, given by Hopfield coefficients described above, depend on the particular shape of the dispersion $E(k)$.

The finite lifetime of exciton-polaritons plays an important role in their condensation, which can be an out-of-equilibrium process. The dynamical nature of condensation becomes especially important in the systems with reduced dimensions, such as quantum wires and pillars. While in planar 2D cavities with sufficiently large spot under favorable conditions polaritons can exhibit a thermal distribution function \cite{Wertz2009}, in 1D cavities with localized pumping an important interplay appears between propagation and relaxation\cite{Wertz2012}. A repulsive potential is created by the reservoir excitons. Polaritons are accelerated and propagate away from the injection spot. This decreases the overlap between wavefunction and the reservoir. This overlap was shown to play a very strong role in polariton condesation, determining the choice of the condensate quantum state \cite{Galbiati2012}.

In periodic systems, the complex structure of the wavefunction and dispersion opens another possibility, that of condensing in a localized excited state, which adds to the options of localized ground state and delocalized (propagative) excited state. Indeed, the repulsive potential of the reservoir, which leads to the delocalization of polaritons by accelerating them away from the pump, becomes in fact attractive for all negative-mass states, that is, roughly, for the upper half of any Brillouin zone. Such states are so preferential for condensation in periodic lattices, that it becomes difficult to avoid them even by using large pumping spots in 2D \cite{Tanese,Jacqmin2014}. In general, the choice of the state for condensation is given by the balance between the scattering into the states and the scattering out of the states, including the decay of polaritons because of the finite lifetime, which therefore becomes an essential parameter, together with the overlap integral. 

\begin{figure}[h!]
\begin{center}
\includegraphics[width=1\textwidth]{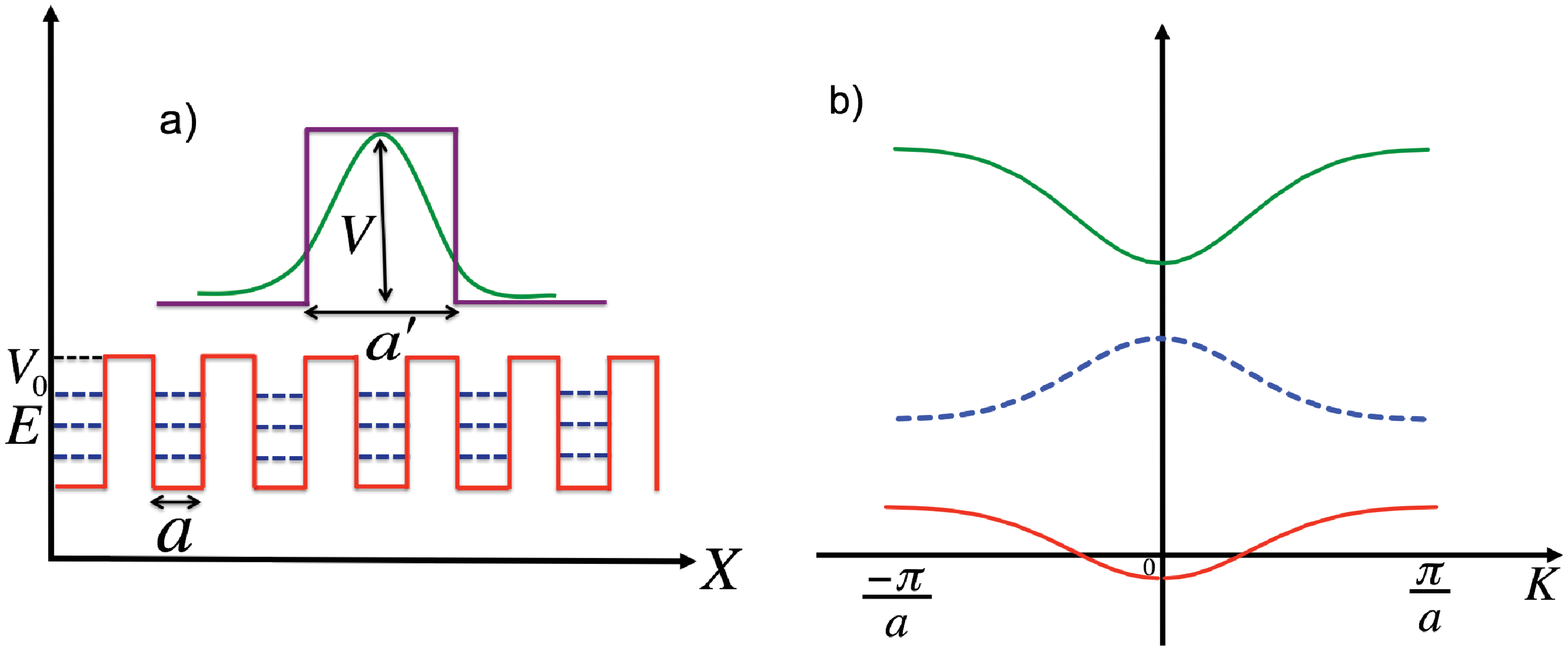}
\end{center}
 \caption{\label{f:scheme}
(Color online) a) Schematic representation of a 1D periodic lattice potential (red), quantized states (blue dashed), reservoir potential (green) and its effective description (violet). b)
The dispersion of the bands of the periodic lattice versus in-plane wave vector (right panel). Each band is formed from a single quantized level.
}
 \end{figure}

The configuration that we study analytically in this work is shown in Fig.\ref{f:scheme}. Polaritons in a periodically modulated wire of average width $d$ are subject to a periodic square potential of height $V_0$ and site width $a$ (period $2a$). We take into account only the lowest state of the \emph{transverse} quantization. Each lattice site is sufficiently deep to contain several longitudinally quantized states for one transverse, requiring $a>d$.  Excitons are created by optical pumping with a Gaussian profile. Because of the repulsive exciton-exciton interactions, they create a repulsive potential for polaritons (green), characterized by its width $a^{\prime}$ and height $V$.

\begin{figure}[tbp] 
\includegraphics[width=0.49\columnwidth]{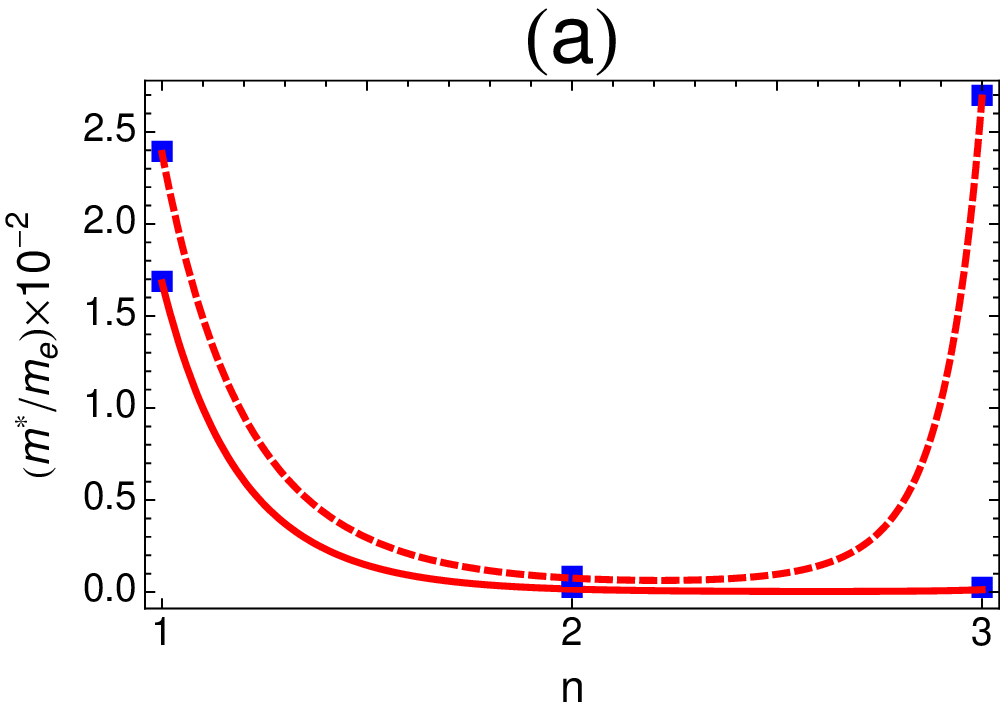}
\includegraphics[width=0.49\columnwidth]{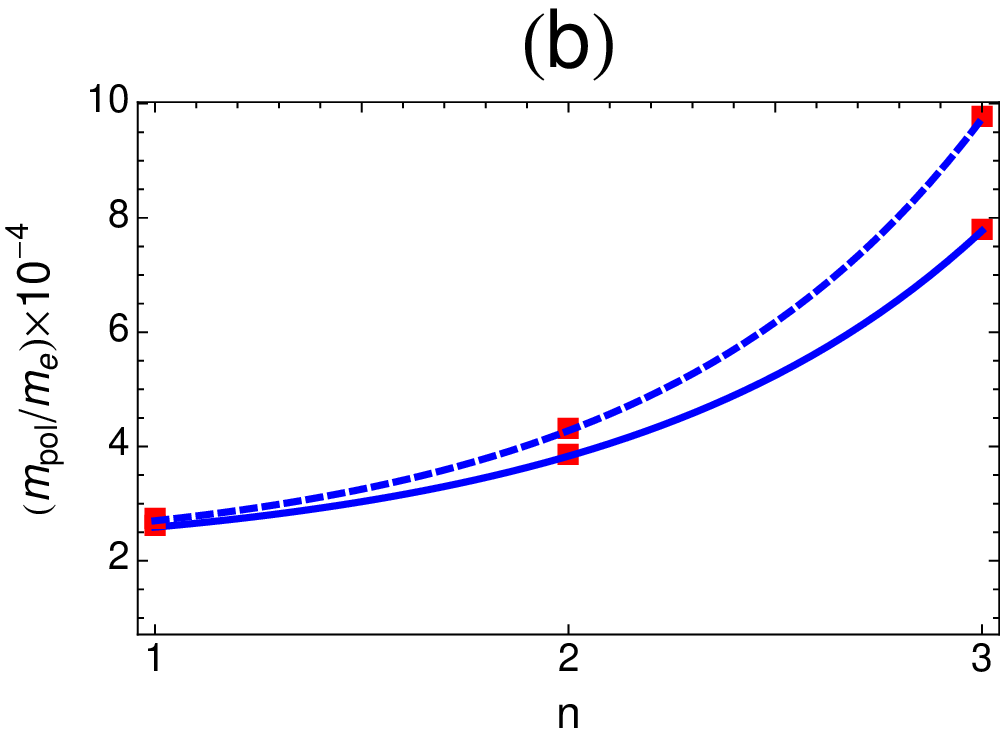}
\caption{(Color online) a) Effective mass $\mathrm{m^*}$ in units of $m_e$ and b) polariton mass $\mathrm{m_{pol}(m_{e})}$, as a function of band number $n$. Solid lines correspond to $\hbar\Omega=15$ meV, and dashed lines to $\hbar\Omega=12$ meV.} 
\label{f:F11}
\end{figure}

Once the potential and the properties of the bare particles are defined, we can calculate the dispersion. We use the tight-binding approximation, requiring the width of each band, proportional to the site-to-site hopping probability, to be considerably smaller than the distance between the on-site quantized levels. Each quantized level $n$ (this level number will be the main argument for plotting all functions below) is determined by the quantization wavevector $k_n=n\pi/a$, which determines the corresponding polariton mass 

\begin{equation}
\mathrm{m_{pol}}=\frac{\mathrm{m_{c}}\mathrm{m_{exc}}}{|C_{k}|^2\mathrm{m_{exc}}+|X_{k}|^2\mathrm{m_{c}}}
\end{equation}
and the quantization energy in a single site $E_n=n^2\pi^2\hbar^2/2m_{pol}(k_n)a^2$. Once these are known, we proceed with the calculation of tunneling probability $T(E_n)$ for a particle with mass $m_{pol}$ and energy of the quantized state $E_n$, which can be easily found analytically for a rectangular potential: 
\begin{equation}
T(E_n)=\frac{4E_n(V_0-E_n)}{4E_n(V_0-E_n)+V_{0}^2\sinh^{2}(\kappa a )},
\end{equation}
 The characteristic decay length under barrier is $\kappa=\sqrt{2m_{pol}(V_{0}-E_n)/\hbar^2}$. We see that the tunneling coefficient, determining the width of a given dispersion band, depends on the level number $n$ via the energy $E_n$ and via the polariton mass $m_{pol}(k_n)$. Once this coefficient is known, one can immediately calculate the dispersion in the nearest-neighbor approximation:

 \begin{equation}
 E^{(n)}(k)=2T(E_n)E_n\left(1-\cos(2k a)\right),
 \end{equation}
The symmetric cosine shape of the dispersion $E_n(k)$ is due to the nearest-neighbor approximation. It allows to determine the effective mass at the extrema of the dispersion in the center and at the edges of the Brillouin zone, which differ only by their signs:
\begin{equation}
\left|m^{*}_n\right|=\hbar^2\left(\frac{\partial^2E^{(n)}(k)}{\partial k^2}\right)^{-1}=\frac{\hbar^2}{8T(E_n)  E_n  a^2}
\end{equation}
The energy dependence of the effective mass is quite complex, because of the polariton mass $m_{pol}(k_n)$ which enters the expression. In Fig. (\ref{f:F11}) we show the result of the calculation of the effective mass (left panel)  and the corresponding polariton mass (right panel) as functions of the quantum number $n$ for two different Rabi splittings $\Omega$. Although $n$ has to be a positive integer, treating it as a continuous variable is useful for analytics, that's why we show continuous dependences together with square points. We have taken the cavity photon mass $\mathrm{m_{c}}=5\times 10^{-5} \mathrm{m_{e}}$, the effective exciton mass $\mathrm{m_{exc}}=0.1\mathrm{m_{e}}$, ($\mathrm{m_{e}}$ is the free electron mass), site width $a=3\mu m$, barrier height $V=0.5\mathrm{meV}$, and detuning $\eta=1\mathrm{meV}$. Both masses are plotted in units of a free electron mass $m_e$. While in the whole range the effective mass is much smaller than $m_0$ (1-2\% of $m_0$), it is at the same time significantly larger than $m_{pol}$ due to the small tunneling probability (narrow bands). The solid line shows the calculation for Rabi splitting of $\hbar\Omega=15\mathrm{meV}$. In spite of the increase of $m_{pol}$, the effective mass $m^{*}$ continues to decrease up to $n=3$ because of the exponential increase of tunneling with energy. However, for a smaller Rabi splitting of $\Omega=12~\mathrm{meV}$, the effective mass (plotted as a dashed line) starts to increase back already for $n=3$, because the non-parabolicity of the polariton dispersion starts to play role for smaller energies.

Now that we have completely determined the properties of quasiparticles for each band $n$, that is, the lifetime and the effective mass, we can forget the underlying structure and consider the quasiparticle in the dispersion as a single object, with its positive or negative effective mass, depending on the band parity and extremum type. In general, we are interested only by negative mass states, which are characterized by localization under repulsive potentials. The lowest defect state appearing due to this potential has a quantization wavevector $k_d=\pi/a'$ and energy $E_d=\pi^2\hbar^2/2m^{*}a'^2$.
The localization length (typical size of the defect state) can be found as a sum of the potential width $a'$ and the decaying exponential tails under the barriers, given by $\kappa(n)=\sqrt{2m^{*}(n)(V-E_d)/\hbar^2}$. The overlap integral $I(n)$ in this case can be estimated as
\begin{equation}
I(n)=\frac{a'}{a'+2/\kappa(n)}
\end{equation}
This overlap integral depends mostly on the effective mass: the smaller the absolute value of the mass, the smaller is also the overlap integral, because the particle spreads away from the defect potential. This leads to the decrease of the particle injection efficiency for this state, because the reservoir acts as a source of the particles for the condensation.

\section{Results and Discussion}

\begin{figure}[h!] 
\includegraphics[width=0.49\columnwidth]{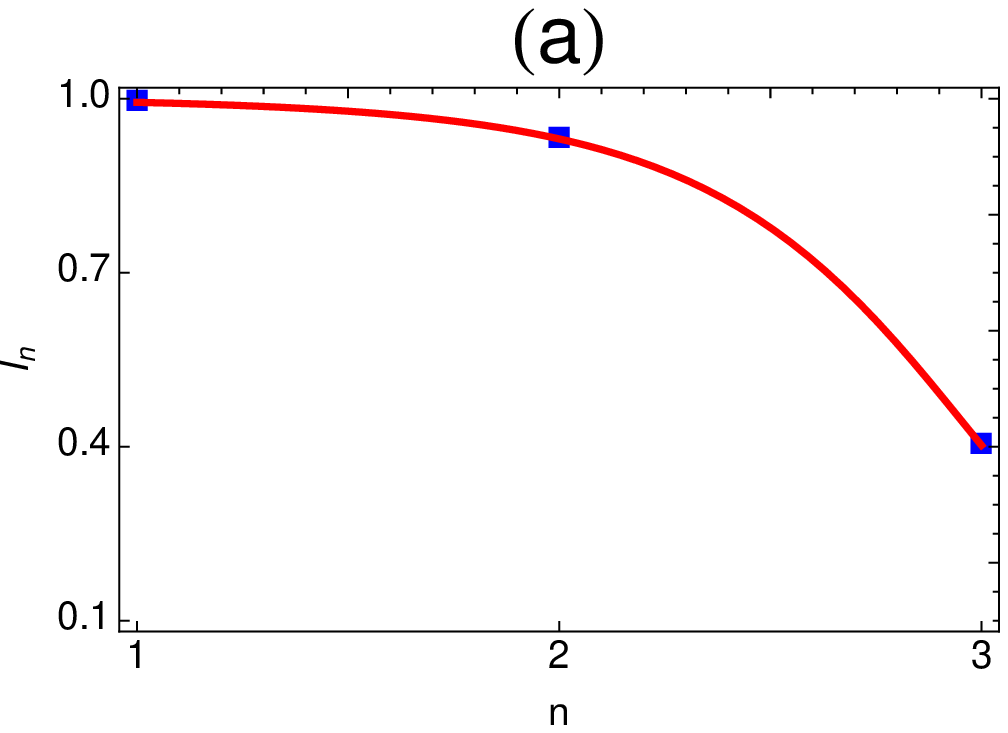}
\includegraphics[width=0.49\columnwidth]{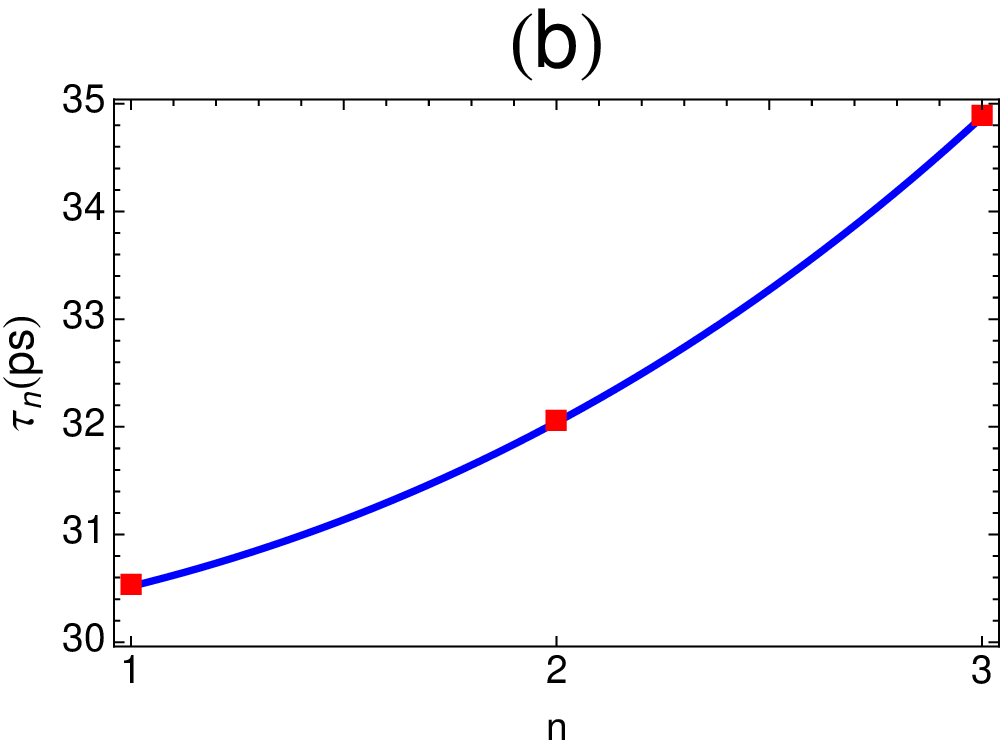}
\caption{(Color online) a) The overlap integral $I_{n}$ and b) the polariton lifetime as functions of the band number $n$ } 
\label{f:F12}
\end{figure}

The choice of the gap with the defect state for the formation of the condensate is finally determined by two crucial parameters: the overlap integral and the lifetime. We plot the dependence of both of them on the band number $n$ in Fig. (\ref{f:F12}): panel (a) shows the overlap integral $I(n)$, which decreases with $n$ (as particles in the bands become lighter) and panel (b) shows the lifetime $\tau(n)$, which increases with $n$ (as polaritons become more excitonic and live longer). We have taken the effective reservoir width $a^{\prime}=0.8\mu\mathrm{m}$, while the other parameters $V$, $\mathrm{m_c}$, $\mathrm{m_{exc}}$, $\Omega$, $\eta$ were kept the same as in Fig. (\ref{f:F11}). Since the number of particles in defect state of a given band is roughly given by the product of these two parameters, we can expect the condensation to occur at the band where the conditions will be optimal.

To describe polariton condensation on the defect states appearing because of the reservoir in periodic polariton lattices we use the Boltzmann equations written for the quantum states found as a solution of the Schrodinger equation. Since the state for condensation is determined by the population below the condensation threshold, we can neglect the modification of the overlap integrals due to the polariton-polariton interactions within the localized states, and also the stimulated scattering terms, which were both important in other works. To capture the essence of the underlying physics, we first consider the case of zero temperature, which greatly  simplifies the Boltzmann equation for the populations $N_n$ of the defect states in the bandgap number $n$: 

\begin{equation}\label{eq:Boltzmann1}
\frac{dN_{n}}{dt}=W_{0}I_{n}N_{R}-\frac{N_{n}}{\tau_{n}}
\end{equation}
Here, $W_{0}$ is the scattering rate of the exciton-polaritons, $I_{n}$ are the overlap integrals between the polariton states and the reservoir, $N_{R}$ is the reservoir population, and $\tau_{n}$ is the polariton lifetime. We neglect the scattering between the localized states which can occur due to polariton-phonon interactions, because these states correspond to different bands, and so their separation in energy is significant, which reduces the efficiency of polariton-phonon interaction. In presence of a strongly populated reservoir, the main mechanism for scattering is the exciton-exciton interaction, resulting in one exciton scattering down from the reservoir into a localized polariton state, and another exciton scattering up within the reservoir, and then quickly relaxing down by exciton-phonon interactions within the reservoir, which are assumed to be very fast. The populaton of each localized defect state is thus given by a balance between the scattering into the state and the lifetime of this state. The polariton distribution in the steady-state regime will be given by the condition $\frac{dN_{n}}{dt}=0$. The corresponding populations are easy to find: 

 \begin{equation}
 N_{n}=W_{0}I_{n}N_{R}\tau_{n}.
 \end{equation}

\begin{figure} 
\includegraphics[width=0.49\columnwidth]{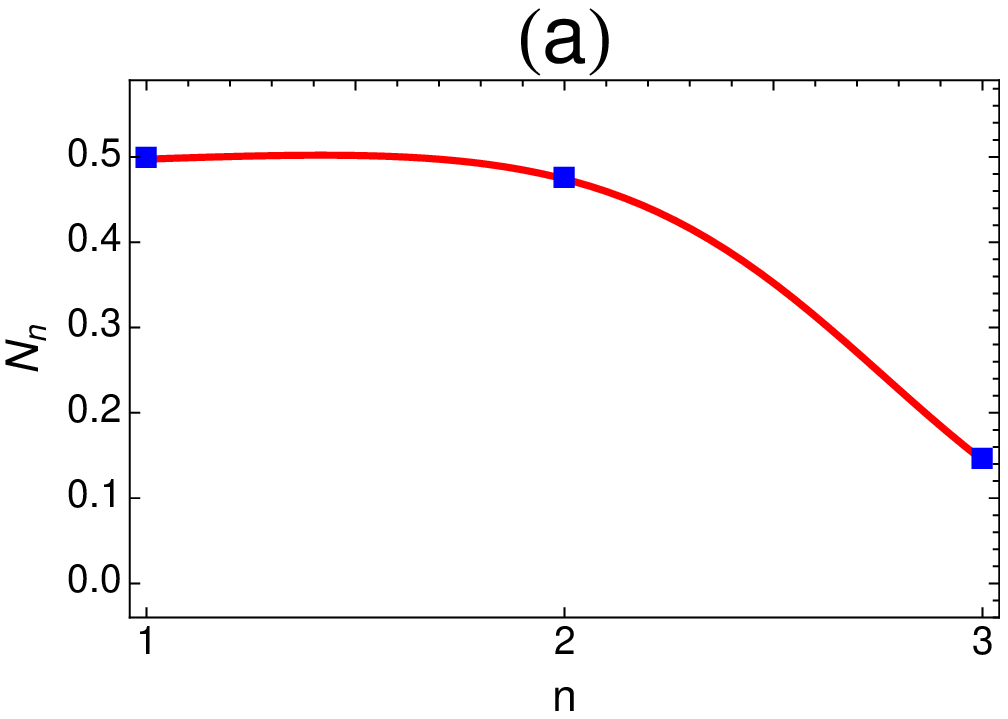}
\includegraphics[width=0.49\columnwidth]{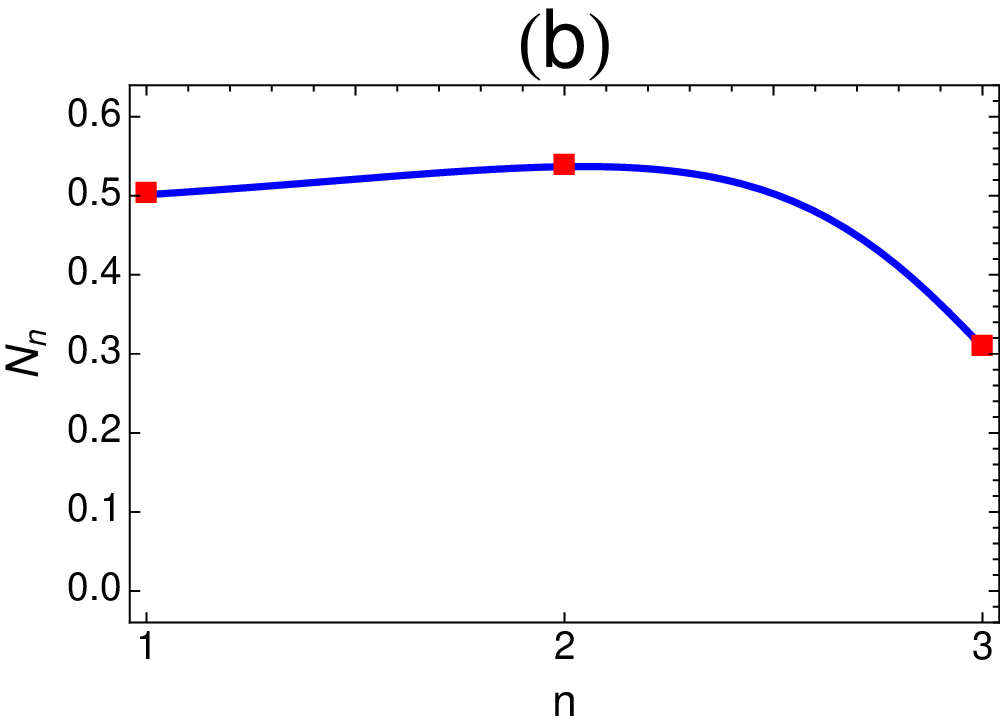}
\caption{(Color online) The polariton occupation numbers for the gap states $N_{n}$, calculated using a semi-classical Boltzmann equation, as a function of the band number $n$. Panels (a) and (b) correspond to different exciton-photon detuning (see text).} 
\label{f:F13}
\end{figure}

As expected, in this simplified picture, the population of a state is proportional to the product of the overlap integral $I_n$ and the lifetime $\tau_n$. The resulting curve is plotted in Fig. \ref{f:F13}. The parameters for the calculation are the potential height $V=0.5\mathrm{m eV}$, the effective cavity-photon mass $m_{c}=0.5 \mu \mathrm{m_{e}}$, $W_{0}=10^5 s^{-1}$, $N_{R}=3.6\times 10^{3}$, the detuning $\eta= 1\mathrm{meV}$ for the right panel and $\eta= 0.065\mathrm{meV}$ for the left panel, $\Omega=15\mathrm{meV}$ and the cavity lifetime $\tau_{c}=30\mathrm{ps}$ and the exciton lifetime, $\tau_{exc}=6\mathrm{ns}$. The scattering rate was chosen small, in order to remain well below the condensation threshold. The polariton population reaches a maximum for $n=2$ for the right panel and $n=1$ for the left panel. So, depending on the parameters, condensation will occur (when the reservoir density is increased) either on the gap state of the 2nd band or on the gap state of the 1st band. The parameters that have the most influence on the choice of this state will be discussed below.

We can now consider a more complicated case and write the Boltzmann equations when the reservoir is at a given non-zero temperature $T$. The equations now contain also the thermal escape of polaritons from the confined defect states into the reservoir by the same mechanism of exciton-exciton scattering. In the steady state with $dN_{n}/dt=0$, the equations read:

\begin{equation}
0=W_{0}I_{n}N_{R}-\frac{N_n}{\tau_{n}}-W_{0}N_{n}e^{\Delta/\alpha}e^{E_{n}/\alpha}.
\end{equation}
 Here, $\Delta=\frac{1}{2}\left(\eta-\sqrt{\eta^2+\Omega^2}\right)$ is the polariton dispersion depth, calculated from the exciton-photon detuning $\eta=E_{x}-\hbar\omega_{0}$ and the exciton-photon coupling  $\Omega$; the temperature is given by $\alpha=k_{B}T$. The equations for different confined states are still not coupled with each other because we neglect the phonon-assisted scattering between the states. Linearizing the exponential term  $e^{E_{n}/\alpha}$ for temperatures higher than the splitting between the bands but lower than the Rabi splitting (this linearization is of course not required here, but greately simplifies the determination of the optimal parameters, discussed in the following), we obtain the approximate analytical solution shown below:

\begin{equation}
N_{n}=\frac{W_{0}\tau_{n}I_{n}N_{R}}{W_{0}\tau_{n}e^{\Delta/\alpha}(1+\frac{E_{n}}{\alpha})+1}.
\end{equation}

\begin{figure}[tbp] 
\includegraphics[width=0.49\columnwidth]{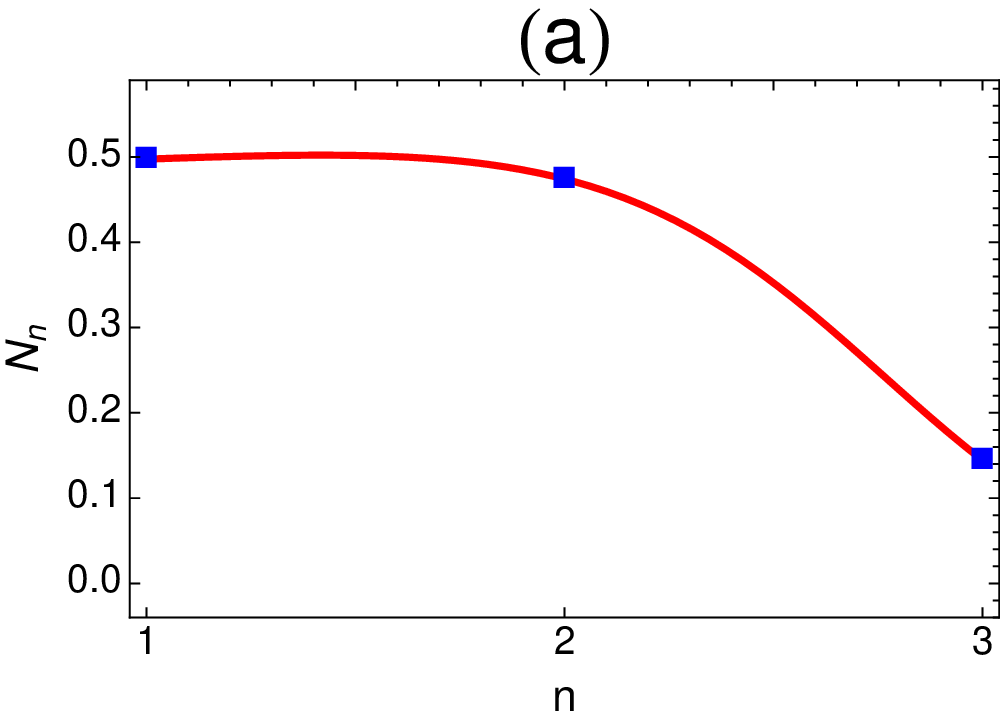}
\includegraphics[width=0.49\columnwidth]{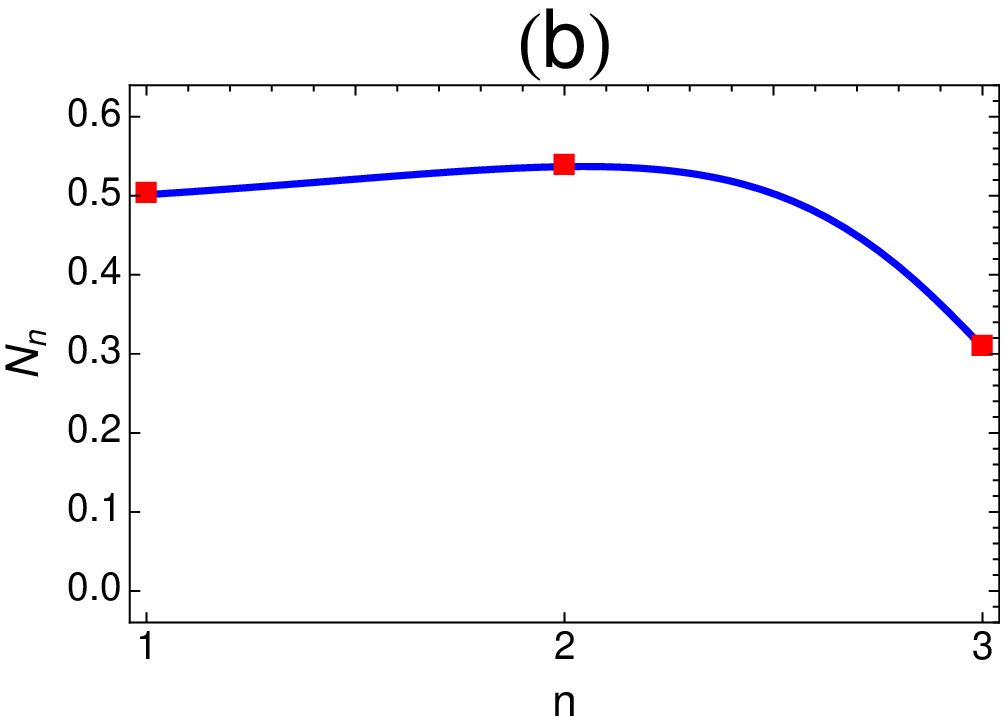}
\caption{(Color online) The polariton occupation numbers for the gap states $N_{n}$, 
as a function of the band number $n$ with the reservoir temperature 10 K 
($\alpha=0.86\mathrm{meV}$). Panels (a) and (b) correspond to different exciton-photon detuning (see text).}
\label{f:F14}
\end{figure}

Figure \ref{f:F14} plots the typical dependence of the polariton occupation number on the band number $n$. The polariton occupation reaches a maximum for $n=2$ for the right panel and $n=1$ for the left panel. 

What are the real system parameters, which determine the optimal conditions for the choice of the defect state for condensation? This is not immediately obvious from the cumbersome expressions for the overlap integral and the polariton lifetime. To answer this question, we have studied all parameters independently, plotting the position of the maximum of $N_n$ as a function of the corresponding parameter. To determine the position of this maximum, we have applied the condition for the derivative of the polariton populations $\partial N_n/\partial n=0$, which is possible if one treats $n$ as a continuous variable.

The strongest dependence was found on the exciton-photon detuning $\eta$ and on the reservoir spatial size $a'$. Both parameters are relatively easy to vary experimentally: the detuning depends on the position on the sample due to the thickness gradient obtained by blocking the substrate rotation in MBE, and the size of the reservoir is controlled by focusing the excitation laser. We show that for both $T=0$ and $T=10$ K the overall behavior is the same, and therefore the simple analytical model at $T=0$ can be used for a quick estimate of the number of the band on which the condensation can be expected to occur.

\begin{figure}[tbp] 
\includegraphics[width=0.49\columnwidth]{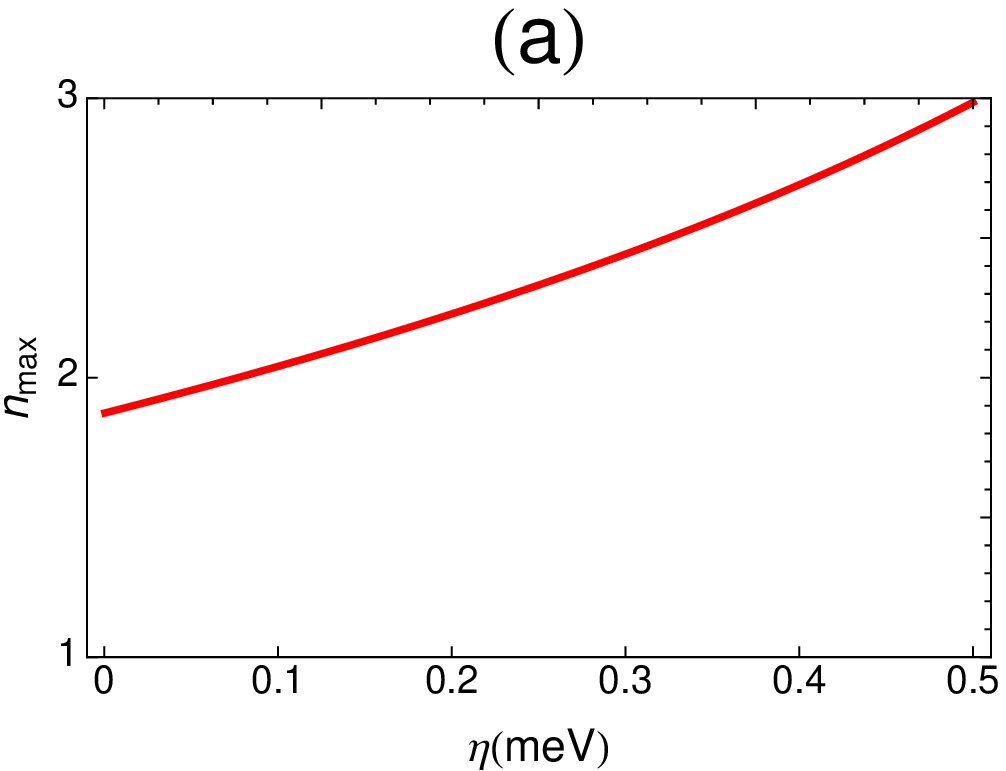}
\includegraphics[width=0.49\columnwidth]{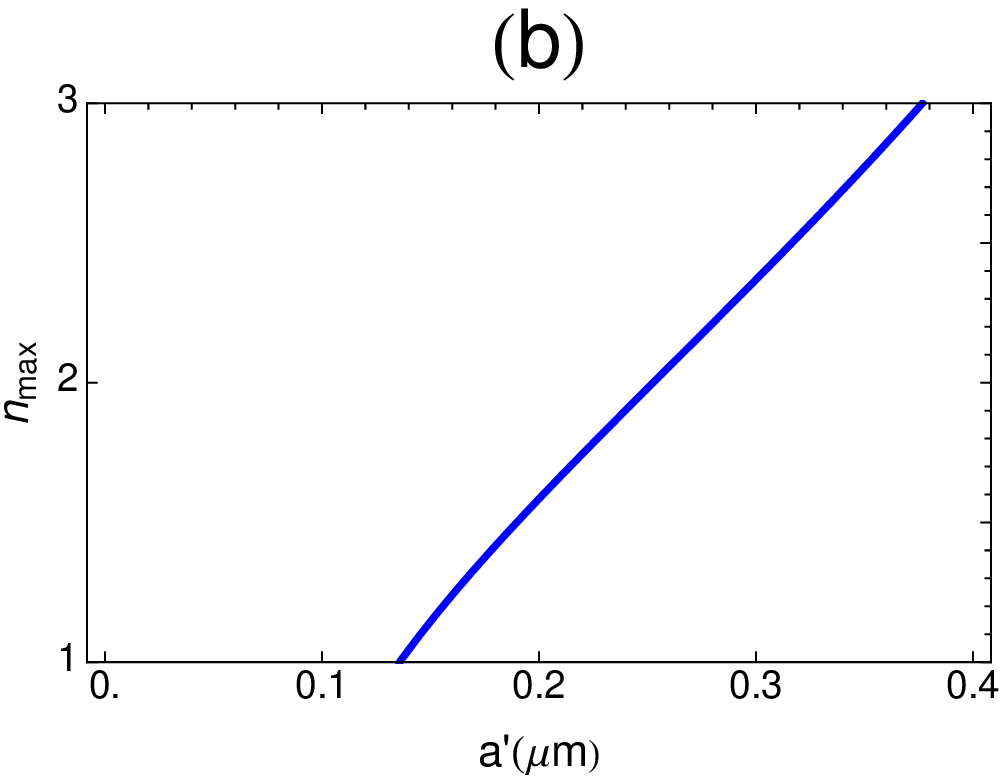}
\caption{(Color online) The position of the maximum population $n_{max}$ 
(a) as a function of the detuning $\eta$ and (b) as a function of the reservoir width $a^{\prime}$ at $T=0$ K.}
\label{f:F21}
\end{figure}

\begin{figure}[tbp] 
\includegraphics[width=0.49\columnwidth]{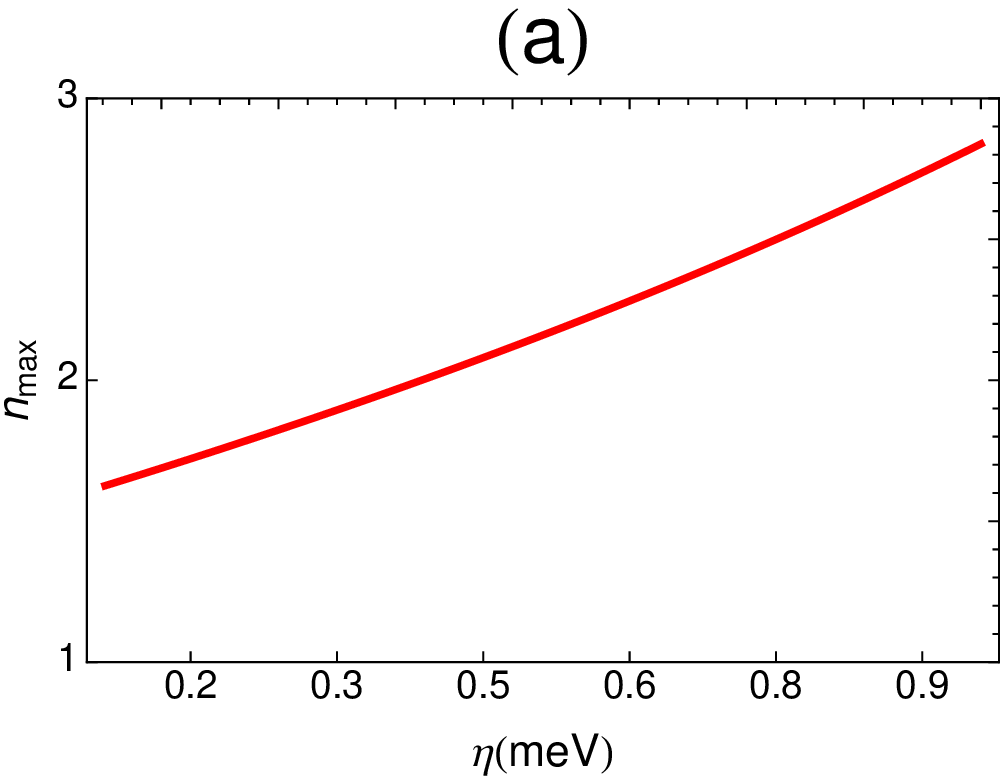}
\includegraphics[width=0.49\columnwidth]{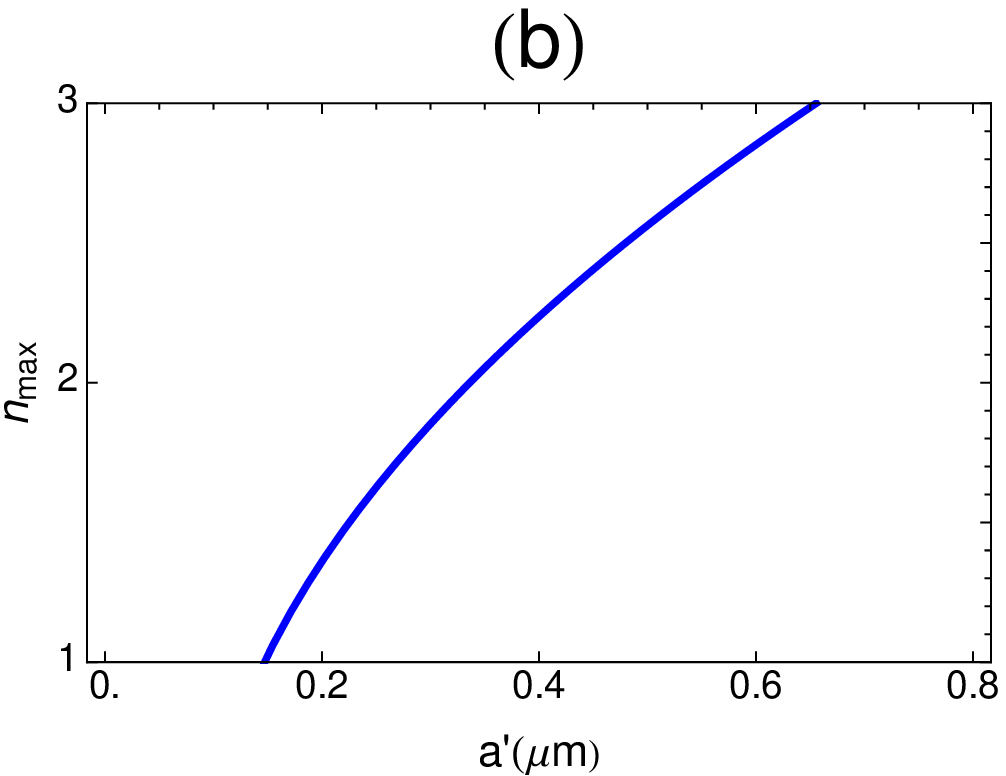}
\caption{(Color online) (Color online) The position of the maximum population $n_{max}$ 
(a) as a function of the detuning $\eta$ and (b) as a function of the reservoir width $a^{\prime}$ at $T=10$ K.} 
\label{f:F31}
\end{figure}

In the left panel of Fig. \ref{f:F21}, we show that $n_{\mathrm{max}}$ increases with detuning $\eta$. On the right panel, $n_{\mathrm{max}}$ also increases with the reservoir width $a^{\prime}$. Both panels of Fig. \ref{f:F31} demonstrate essentially the same behavior. It should be noted that such narrow reservoir potential $a'$ can be essentially considered as a Dirac delta function within the effective mass approximation. However, its height should not be too important, otherwise it can simply break down the approximation. Thus, we can conclude that varying the easily accessible experimental parameters one can switch the condensation between states in different bandgaps. This result, bearing certain similarities with the recent proposal of polariton router \cite{Savenko2013,Marsault2015}, can be useful for all-optical on-chip information treatment systems.

Finally, condensation in the true ground state in such system could be expected in the case of a very large pumping spot, whose size can be estimated as $a'=v_{g}\tau_{1}=4aE_1T_1/\hbar$. In this case, the overlap integral is not determined by the particle mass, but rather by the lifetime, which becomes a limiting factor. However, even if the condensation occurs at $k=0$ under the pump, polaritons will still be accelerated outside of the spot, giving the impression of non-equilibrium condensation.

\section{Conclusions}

To conclude, we have analytically studied the polariton condensation on gap states in presence of a localized non-resonant pumping. We find that the choice of the state for condensation is determined by the balance between the overlap integral and the lifetime, and that it can be controlled by varying detuning and the reservoir size (laser spot size). With realistic parameters, condensation can occur either on the gap state of the lowest band, or on the gap state of the 2nd band. This approach gives a general understanding of the results of recent experiments on gap solitons.

We acknowledge the support of ANR GANEX (207681) and Quandyde (ANR-11-BS10-001) projects.

\bibliography{reference}

\end{document}